# Interconnected Linguistic Architecture

Johannes Härtel, Lukas Härtel, Marcel Heinz, Ralf Lämmel, and Andrei Varanovich[a]

a   Software Languages Team, Faculty of Computer Science, University of Koblenz-Landau, Germany

**Abstract**    The *context* of the reported research is the documentation of software technologies such as object/relational mappers, web-application frameworks, or code generators. We assume that documentation should model a macroscopic view on usage scenarios of technologies in terms of involved artifacts, leveraged software languages, transformation and conformance relationships, I/O behavior, and others. In previous work, we referred to such documentation also as 'linguistic architecture'. The corresponding models may also be referred to as 'megamodels' while adopting this term from the technological space of modeling/model-driven engineering.

   This work is an *inquiry* into making such documentation more useful by means of connecting (mega)models, systems, and developer experience in several ways.

   To this end, we adopt an *approach* that is primarily based on prototyping (i.e., implementation of a megamodeling infrastructure with all conceivable connections) and experimentation with showcases (i.e., documentation of concrete software technologies).

   The *knowledge* gained by this research is a notion of interconnected linguistic architecture on the grounds of connecting primary model elements, inferred model elements, static and runtime system artifacts, traceability links, system contexts, knowledge resources, plugged interpretations of model elements, and IDE views. A corresponding suite of aspects of interconnected linguistic architecture is systematically described.

   As to the *grounding* of this research, we describe a literature survey which tracks scattered occurrences and thus demonstrates the relevance of the identified aspects of interconnected linguistic architecture. Further, we describe the `MegaL/Xtext+IDE` infrastructure which realizes interconnected linguistic architecture.

   The *importance* of this work lies in providing relatively formal, ontologically rich, navigable, and verifiable documentation of software technologies, thereby helping developers to better understand how to use technologies in new systems (prescriptive mode) or how technologies are used in existing systems (descriptive mode).

# The Art, Science, and Engineering of Programming

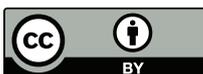

**Interconnected Linguistic Architecture**

## 1 Introduction

**Research context: *software architecture meets technology documentation*** Software architecture may be concerned with a software system's physical and logical composition, data layout, requirements, interfaces, and other aspects. In this paper, we focus on another notion of architecture, *linguistic* architecture [10, 20], which is concerned with usage scenarios of software technologies at a more macroscopic level: What are the artifacts involved in technology usage? What software languages are leveraged? How are artifacts related by conformance relationships or otherwise? What technology-related I/O behavior is exercised, e.g., through transformation, generation, or serialization? The corresponding models may also be referred to as 'megamodels' while adopting this term from the technological space of modeling/model-driven engineering [4, 3]. Accordingly, modeling linguistic architecture may also be referred to simply as 'megamodeling'.

Models of linguistic architecture (i.e., megamodels) may serve the documentation of software technologies. Such models are assumed to complement classical documentation because of the applied ontological discipline. In this paper, we use the megamodeling language MegaL.[1] (We recap MegaL in Sec. 2.) In this paper, we pick the Eclipse Modeling Framework[2] (EMF) as the running example of a software technology to be documented. EMF is a core technology in the technological space of modeling/model-driven engineering (MDE). The approach of this paper has also been applied to, for example, XML data binding, Object/Relational mapping, and web-application frameworks. The megamodel fragment in Fig. 1 models one aspect of (using) EMF: the generation of a EMF Model API from a metamodel (*metaModel*) and a generator model (*genModel*); the generation is abstracted as a function (*EMFGenerator*); the generated API uses just a subset of Java (*EcoreJava*).

```
...
EcoreJava subsetOf Java // An EMF Model API is valid Java
genModel elementOf EMFGenModel // Generator configuration
genModel references metaModel // Reference to the metamodel
javaFiles elementOf EcoreJava // The modeled/defined API
EMFGenerator(genModel) ↦ javaFiles // Application of generator
...
```

■ **Figure 1** A megamodel fragment documenting generation of an EMF Model API.

Without going into details of the MegaL language here, the example conveys that a model of linguistic architecture involves *entities* such as languages, artifacts, and functions; further, the megamodel states *relationships* between the entities such as 'membership' (or 'elementOf' to assign languages to artifacts) or 'mapping' (by means of function application). One may say that megamodels capture an ontology of software technologies. Thus, a language like MegaL may be said to be both a domain-

---

[1] http://www.softlang.org/megal
[2] https://eclipse.org/modeling/emf/

Johannes Härtel, Lukas Härtel, Marcel Heinz, Ralf Lämmel, and Andrei Varanovich

specific modeling language (with linguistic architecture or technology documentation as domain) and a knowledge representation language.

**Research objective:** *making megamodels more useful*  Models of linguistic architecture, in order to be useful as documentation, should properly help developers to better understand how to use technologies in new systems (prescriptive mode) or how technologies are used in existing systems (descriptive mode). Megamodels, in general and, in the case of models of linguistic architecture, in particular, reside at a high level of abstraction. Thus, the key question is this: How to make megamodels useful enough for creating or understanding systems, as far as technology usage is concerned? A subordinated question is this: How to exactly perform megamodeling? Consider Fig 2 for preparing for a deeper discussion of the use of linguistic architecture in software development.

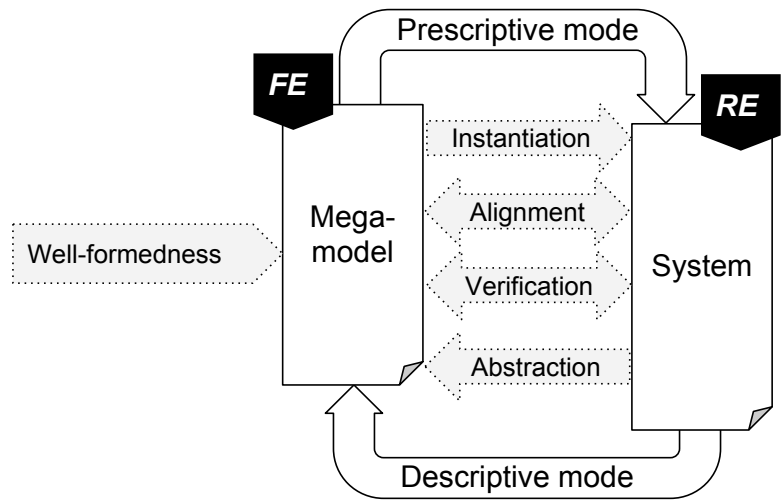

**Figure 2**  Linguistic architecture in forward and reverse engineering (FE & RE).

In the case of *forward engineering* (FE), we start from a megamodel for scenarios of technology usage which we *instantiate* (manually or semi-automatically) to derive the relevant parts of the system; this is the prescriptive mode of megamodeling. In the case of *reverse engineering* (RE), we start from a system from which we *abstract* (manually or semi-automatically) to derive a megamodel; this is the descriptive mode of megamodeling. In fact, both model and system may (partially) co-exist right from the start and thus, they need to be *aligned* (manually or semi-automatically). In all cases, the megamodel, just by itself, must be *well-formed*, i.e., it should make correct use of the megamodeling vocabulary. More importantly, model and system should be consistent with each other, i.e., model elements should be linked to suitable system artifacts and relationships on model elements should somehow correspond to actual properties in the system. To this end, we assume a form of *verification*. In the descriptive mode, successful verification would mean that the derived model is a correct abstraction of the system. In the prescriptive mode, successful verification would mean that the derived system is complete and consistent with regard to the model.

In returning to the key question stated above, in order to make megamodels useful enough, developer support is needed for instantiation, abstraction, alignment, and verification. The paper's contribution aims at such support in a comprehensive manner. (Checking well-formedness of megamodels is a relatively mundane task.)

**Contribution of the paper: *Interconnection*** In previous work [20], we introduced basic alignment and verification techniques. That is, entities of the megamodel can be bound to actual artifacts or knowledge resources; also, relationships can be interpreted on artifacts in the actual system. In the present paper, we describe a general and comprehensive approach towards connecting (mega)models, systems, and developer experience, thereby supporting instantiation, abstraction, alignment, and verification, as discussed above. We identify several aspects of connecting primary model elements, inferred model elements, static and runtime system artifacts, traceability links, system contexts, external knowledge resources, plugged interpretations of model elements, and IDE views, as summarized in Fig. 3.

> **Artifact binding** Links to artifacts in a system are part of the model; they are attached to corresponding entities in the model; different kinds of destinations are handled by plugins; the URI format is used.
>
> **Semantic annotations** Models are annotated in the sense of the Semantic Web; model elements are associated with objects from knowledge resources such as schema.org, DBpedia, or Wikipedia.
>
> **Pluggable analyses** Relationships between entities are subject to pluggable analyses that give feedback to the user; the functionality may be readily available or require designated effort.
>
> **Modularized models** Each module corresponds to a context of language and technology usage; reuse corresponds to the formation of larger (compound) contexts.
>
> **Transient artifacts** These artifacts are accessed through some means of interception, as they manifest temporarily as the result of specific actions performed by or with the system.
>
> **Model inference** There exist implied model elements, e.g., fragments of artifacts due to parthood; these model elements are inferred so that they do not need to be modeled explicitly.
>
> **Explorable connections** All 'connections' are explorable by the user; for instance, one may navigate from model elements to system artifacts or relevant plugins for analysis.
>
> **Traceability links** Relationships between artifacts that consist of parts may be represented as collections of traceability links, i.e., bipartite graphs. These links are inferred and amenable to navigation (exploration).

■ **Figure 3** Aspects of interconnected linguistic architecture.

For each aspect, we will describe its purpose and characteristics. The aspects are evaluated by means of a literature study. The resulting notion of interconnected linguistic architecture is realized, to a large extent, in the `MegaL/Xtext+IDE` infrastructure.

Johannes Härtel, Lukas Härtel, Marcel Heinz, Ralf Lämmel, and Andrei Varanovich

**Road-map of the paper** Sec. 2 applies megamodeling to the documentation of technologies; the aspects of interconnected linguistic architecture are motivated along the way. Sec. 3 develops the aspects of Fig. 3 and discusses their realization in the MegaL/Xtext+IDE infrastructure. Sec. 4 evaluates our development in terms of both a literature study on interconnected linguistic architecture and the status of our realization. Sec. 5 discusses related work on top of the aforementioned literature study. Sec. 6 concludes the paper.

## 2 Documentation of technologies

In this section, we demonstrate the megamodeling-based approach to technology documentation. We focus on the basic tenet of megamodeling and corresponding language constructs of MegaL. Aspects of interconnected linguistic architecture are motivated along the way. The proper description and illustration of the aspects are provided in Sec. 3.

### 2.1 Stories on linguistic architecture

Our demonstration is concerned with modeling a domain-specific language (DSL) with the Eclipse Modeling Framework (EMF) and Xtext.[3] A series of scenarios is presented; the scenarios build on top of each other. We describe the scenarios in a story format. Our descriptions proxy for informal documentation. We prepare a process of inquiry so that megamodel elements are extracted from the stories.

**XML:** XML is a data exchange format and XSD is a schema language for XML documents. XSD schemas are written in XML.

**EMF:** Ecore is another format to define a schema. Here, schemas are called *metamodels* and documents are called *models*. A model may be persisted as an XMI file. XMI is an XML language. Metamodels are models, too. The language of metamodels (Ecore) is defined by the metametamodel which conforms to itself (i.e., it is self-descriptive).

**ATL:** This is a language for transformations; input and output conform to some metamodels. The transformation is applied on a model that conforms to the input metamodel resulting in a model that conforms to the output metamodel. (We skip over a detail here: the description of the transformation also conforms to another metamodel that is specific to ATL.)

**Xtext:** This technology supports the development of a language workbench based on a grammar. Xtext integrates with EMF modeling; the corresponding metamodel is either defined explicitly or derived from a grammar by the Xtext generator. The derived metamodel does not contain any operations. The Xtext generator is configured in the MWE2 language as a component-based workflow. Running this generation process constructs the derived artifacts in the file system. The rules of an Xtext grammar

---

[3] https://eclipse.org/Xtext/

| Id | Question | Relevant MegaL constructs |
|---|---|---|
| L1 | Which languages can be identified? | Type *Language* |
| L2 | Is one language contained in another? | Relationship *subsetOf* |
| A1 | What artifacts participate in the scenario? | Type *Artifact* |
| A2 | What is the language of each artifact? | Relationship *elementOf* |
| A3 | Does an artifact conform to another artifact? | Relationship *conformsTo* |
| A4 | Does an artifact define a language? | Relationship *defines* |
| F1 | Is one artifact derived from another artifact? | Type *Function* |
| F2 | What is domain and range of a function? | Function with domain & range |
| F3 | How is a function applied? | Function application '$f(x) \mapsto y$' |
| F4 | How is a function defined? | Relationship *defines* |
| R1 | Are artifacts closely similar to each other? | Relationship *correspondsTo* |
| R2 | Can a correspondence be structured? | Relationship *partOf* |
| R3 | What causes a correspondence? | Function application '$f(x) \mapsto y$' |
| C1 | Can the entity be described conceptually? | Type *Concept* |
| C2 | Does the entity use the concept? | Relationship *uses* |
| C3 | Does the entity help to use the concept | Relationship *facilitates* |

■ **Table 1** Discovery of entities and relationships.

corresponds (essentially) to the derived `Ecore` classes. This correspondence depicts the mapping between concrete and abstract syntax.

**EMF Model API:** Interaction with models is supported by a generated metamodel-specific EMF Model API on transient object graphs. Modifications are exchanged between the XMI persistence layer and the JVM representation of models by serialization.

## 2.2 Questions on linguistic architecture

Megamodels involve entities such as languages, technologies,[4] functions, and artifacts. Megamodels also involve relationships such as conformance or correspondence. We submit the questions of Table 1; they are helpful in discovering entities and relationships. We associate questions with language elements of MegaL.

The list of questions essentially covers the declarations in the *prelude*—this is MegaL's base library module written in MegaL itself. The prelude contains the predefined knowledge needed to commence with linguistic modeling. Adding and removing questions can be compared to an evolution of the prelude that is currently part of our research. MegaL supports type specialization for adding new entity subtypes. Thus, adding another subtype to the prelude would presumably trigger the inclusion of another question. For instance, if we were to add a type *QueryLanguage*, then we would also need to include a corresponding question: Can a language be identified that expresses queries on data—presumably without causing side effects?

---

[4] In this paper, for brevity, we do not declare entities of type 'technology'; instead, we may discuss the 'functions' of the technologies.

Johannes Härtel, Lukas Härtel, Marcel Heinz, Ralf Lämmel, and Andrei Varanovich

### 2.3 Models of linguistic architecture

By going through the questions in the table, one may set up the major model elements in stories on linguistic architecture, as captured by the following MegaL modules.

**Megamodel of the XML story**

```
1 module XML import (Prelude) // Import basic vocabulary
2 XML : Language // Declare XML as a language entity
3 XSD : Language // and XSD (XML Schema), too
4 XSD subsetOf XML // Subset relationship on XSD and XML
5 xmlFile : Artifact // Declare artifact
6 xsdFiles : Artifact+ // Declare collection of artifacts
7 xmlFile elementOf XML // Assign language to artifact
8 xsdFiles elementOf XSD // Assign language to artifact
9 xmlFile conformsTo xsdFiles // XSD—based validation
```

MegaL's prelude readily provides an entity type *Language*, as required for an answer of question L1: XML and XSD are declared as entities of type *Language* (lines 2–3; notation for entity declarations: *e* : *t* with *e* the entity's name and *t* its type). Languages are conceived as sets and thus, they may be related by a *subsetOf* relationship (again from the prelude), which is to be applied to XSD and XML (line 4), thereby answering question L2 (notation for relationship declarations: *entity relationshiptype entity*).

The prelude's type *Artifact*, as mentioned in question A1, conveys that an entity is considered an actual resource such as a file in a system: *xmlFile* (line 5) is a proxy for an XML file; *xsdFiles* (line 6) is a proxy for a collection (see the '+') of schema files, as XML schemas may be modularly decomposed into multiple files. Two relationships are at play:

- *elementOf* associates an artifact with a language (question A2). In the XML model, *xmlFile* is written in XML, *xsdFiles* is written in XSD (lines 7–8).
- *conformsTo* associates artifacts; one artifact conforms to the other artifact (question A3). In the XML model, *xmlFile* conforms to *xsdFiles* (line 9).

Let us assume that *xmlFile* and *xsdFiles* proxy for actual artifacts in a system. If we wanted to check whether the declared relationships (about language membership and conformance) actually hold, we require the effective binding of the model elements for the artifacts to actual artifacts and automated tests for membership and conformance. This will be addressed by the aspects 'Artifact binding' and 'Pluggable analyses'.

**Megamodel of the EMF story**

```
1 module EMF import (Prelude)
2 Ecore : Language // As defined by metaMetaModel
3 Custom : Language // As defined by metaModel
4 metaModel : Artifact // A metamodel artifact
5 metaMetaModel : Artifact // The metametamodel
6 metaModel elementOf Ecore
7 metaMetaModel elementOf Ecore
8 metaModel conformsTo metaMetaModel
9 metaMetaModel conformsTo metaMetaModel
10 metaModel defines Custom
```

**Interconnected Linguistic Architecture**

```
11 metaMetaModel defines Ecore
```

When compared to the XML and XSD languages of the XML module, the languages Ecore and *Custom* are characterized in terms of artifacts that actually define them (lines 10–11). We leverage the relationship type *defines* for associating an artifact with the defined language (question A4). In particular, *metaModel* defines a *Custom* language—this is the DSL to be modeled.

**Megamodel of the ATL story**

```
1  module ATL import (EMF) // ATL depends on EMF
2  transformation : Custom → Custom // Function on DSL
3  input : Artifact // Input artifact
4  output : Artifact // Output artifact
5  input elementOf Custom // input (source) of transformation
6  output elementOf Custom // output (target) of transformation
7  transformation(input) ↦ output // Function application
8  ATL : Language // The ATL language
9  atlmodule : Artifact // An ATL transformation module
10 atlmodule elementOf ATL
11 atlmodule defines transformation // Semantics of ATL module
```

That is, we reuse the megamodel for EMF, thereby applying the interconnection aspect 'Modularized models'. We model an endogenous transformation, i.e., source equals target language. Questions F1 and F2 are answered by introducing a function entity named *transformation* (line 2) with *Custom* for its domain and range (notation for function declarations: *function : domain → range*). Question F3 is answered by applying the function on the *Custom* Language elements *input* and *output* (lines 3–7; notation for function application: $f(x) \mapsto y$). There is a *defines* relationship (line 11) between artifact *atlmodule* written in the ATL language (lines 9–10) and *transformation*, thereby answering question F4. (MegaL's prelude overloads the range of *defines* on languages and functions.)

**Megamodel of the Xtext story**

```
1  module Xtext import (EMF) // Xtext integrates with EMF
2  Xtext : Language // Xtext language
3  grammar : Artifact // An artifact for the grammar
4  grammar elementOf Xtext // An Xtext grammar
5  EcoreWithoutOps : Language // Relevant subset of Ecore
6  EcoreWithoutOps subsetOf Ecore
7  metaModel elementOf EcoreWithoutOps // Restriction of import
8  metaModel correspondsTo grammar // Correspondence
9  generator : Xtext → EcoreWithoutOps // Generator function
10 generator(grammar) ↦ metaModel // Generator application
11 MWE2 : Language // Language for generator configuration
12 workflow : Artifact // Workflow artifact
13 workflow elementOf MWE2 // Workflow is written in MWE2
14 workflow defines generator // Workflow defines generator function
```

There is the grammar notation Xtext (line 2) and actual grammar artifact (lines 3–4). We also introduce the Ecore subset *EcoreWithoutOps* (lines 5–6) which is used in combination with Xtext. The imported metamodel is constrained accordingly (line 7).

The grammar is declared to *correspond to* (i.e., to be closely similar to) the metamodel (line 8), thereby answering question R1. The correspondence can be structured in terms of the parts of the corresponding entities such as grammar rules or metamodel classes in the case at hand, thereby answering question R2. The (application of the) *generator* function gives the cause of the correspondence between the artifacts *metaModel* and *grammar* (lines 9–10), thereby answering question R3. Generation is configured with the help of a designated *MWE2* language (lines 11–14).

The aspect 'Traceability links' suggests that the structured correspondence should be modeled by links. The interconnection aspect 'Model inference' suggests that the links should be inferred and the aspect 'Explorable connections' suggests that one can navigate along the links (in an IDE).

**Megamodel of EMF Model API story**

```
module EMFModelAPI import (
  EMF, // The EMF module is enhanced
  XML) // The XML module is needed for serialization
Java : Language // Java is a Language
EcoreJava : Language // A Java subset for EMF Model APIs
EcoreJava subsetOf Java // An EMF Model API is valid Java
EMFGenModel : Language // Language for the generator model
genModel : Artifact // Parameters of the generation
genModel elementOf EMFGenModel
genModel references metaModel // Referencing
EMFGenerator : EMFGenModel → EcoreJava
EMFGenerator(genModel) ↦ api // Application of generator
CustomObjects : Language // Object graphs for Custom
Serialization : CustomObjects → Custom
Deserialization : Custom → CustomObjects
XMI : Language // Format for default persistence for EMF
XMI subsetOf XML // XMI is a subset of XML
Custom subsetOf XMI // Custom uses default persistence
javaFiles : Artifact+ // The modeled/defined API
javaFiles elementOf EcoreJava
metaModel correspondsTo javaFiles // Close resemblance
javaFiles defines CustomObjects
javaFiles defines CustomSerialize
javaFiles defines CustomDeserialize
model : Artifact // A serialized artifact
model elementOf Custom // ... of Custom language
model conformsTo metaModel // Conformance to metamodel
objectGraph : Transient // A runtime artifact
objectGraph elementOf CustomObjects
objectGraph conformsTo javaFiles // Conformance to Java classes
CustomSerialize(objectGraph) ↦ model
CustomDeserialize(model) ↦ objectGraph
```

EMF's generation facility generates a EMF Model API from a metamodel. The generator is modeled by the *EMFGenerator* function which takes a designated *genModel* to produce the *javaFiles* which in turn defines an object-graph language *CustomObjects* for *Custom* and the functions *CustomSerialize* and *CustomDeserialize* for de-/serialization. For instance, deserialization would result in a 'transient' (in-memory) object graph to be treated in a special manner during verification (megamodel versus system); see the interconnection aspect 'Transient artifacts'.

The default EMF generation builds an XMI base persistence mechanism so language *Custom* is now stated as a subset of XMI. (Xtext leverages parser-based persistence instead, but this is not modeled here.) The fact that our model deals with the serialization and persistence concepts is only captured, so far, in this explanatory text. Eventually, we should promote such concepts to proper model elements with appropriate links to knowledge resources, thereby answering questions C1-C3; see the aspect 'Semantic annotations'.

## 3  Aspects of interconnected linguistic architecture

We describe the aspects using the following format:

**Problem**   A problem that modelers or developers face when using megamodels descriptively or prescriptively. (The problem would have to do essentially with mastering/understanding instantiation, abstraction, alignment, or verification, as discussed in the introduction.)

**Solution**   The proposed solution to the problem; the solution may affect the megamodeling language, the encompassing infrastructure, and the methodology for megamodeling.

**Realization**   The realization of the solution in the MegaL language and the MegaL/Xtext+IDE infrastructure.

### 3.1  Artifact binding

**Problem**   A model describes or prescribes, among others, the artifacts in a system subject to modeling. The model would be hard to understand, if it does not provide links to the artifacts of the modeled system.

Megamodels and systems could reside in repositories, be checked out into a local file system, or shared across a network. Links from megamodels to systems necessitate different resolution protocols. Also, artifacts may transcend the naive representation as a file. For instance, artifacts of interest may be fragments of files, e.g., a method declaration in Java. Fig. 4 illustrates some connection forms.

**Solution**   Model elements are bound to artifacts. The range of bindings is, in agreement with the diversity of destinations, selected to be the *Uniform Resource Identifier* notation. As a result, both local files and web resources may be linked. The support for fragment URIs (suffixes that point to anchors inside the destination) allows identifying nested elements. URI resolution is pluggable to deal with the diversity of endpoints. URIs are hierarchically processed to handle cascaded resolution, e.g., when resolution begins in a repository, proceeds with selection of a file resource, further proceeds with parsing to select an AST node as a fragment; see Fig. 5 for illustration.

**Realization**   In MegaL, links can be assigned to model elements like this:

Johannes Härtel, Lukas Härtel, Marcel Heinz, Ralf Lämmel, and Andrei Varanovich

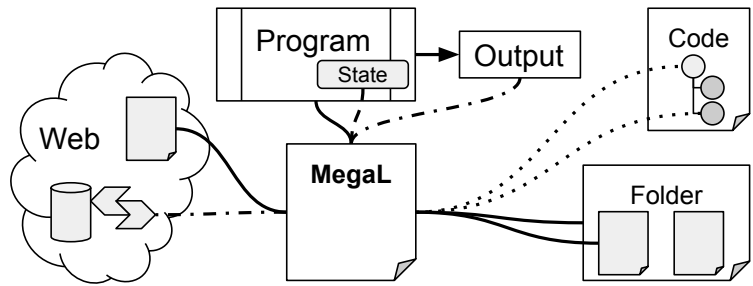

**Figure 4** Interconnection of megamodel and system; solid lines represent direct links; dashed lines indicate introspection; dash-dotted lines show interception; dotted lines depict fragment links.

`github://user/project/files/data.jar/content.xml/root/models/model#1`

**Figure 5** The given URI is handled by three different resolution methods, the first being a GitHub repository resolver, followed by a JAR unpacker, and concluded by an XML parser navigating to a certain element.

```
// module EMF continued
metaMetaModel = 'eclipse:/org.eclipse.emf.ecore/model/Ecore.ecore'
```

Here, we assume support for a destination of 'eclipse plugins'. That is, the metametamodel should be located in a certain directory as part of the identified EMF plugin. MegaL/Xtext+IDE implements bindings as properties of megamodel entities. Bindings can take any form, as they are represented as Java objects, but the textual representation and serialization requires essentially URIs. Resolution is pluggable by implementing an artifact provider interface:

- `accept(x: Object): boolean` The method is true, if the provider can navigate in an object.
- `next(x: Object): [String]` The method returns the URI segments that can be navigated to.
- `navigate(x: Object, key: String): [Object]` The method resolves a key $\in$ next(x) to the actual resource.

In case studies, several providers were implemented, e.g., a Java object provider using reflection, a file-system navigation, a classpath search engine that allows integration of compiled Java classes, and a Java AST traversal provider.

### 3.2 Semantic annotations

**Problem** Names of model elements should be drawn from appropriate vocabularies of the relevant technology and the broader domain of software engineering. A model would be hard to understand, if the correspondence of names to relevant vocabulary would be non-obvious or ambiguous.

**Solution** Model elements are semantically annotated by linking them to appropriate knowledge resources, e.g., pages on Wikipedia, DBpedia, or resources on schema.org. How exactly model elements exactly should be annotated and what sources to use,

**Interconnected Linguistic Architecture**

depends, of course, on the specific megamodeling vocabulary. For instance, one could link languages or technologies to resources.

| XML | http://dbpedia.org/page/XML |
|---|---|
| EMF | https://eclipse.org/modeling/emf/ |

Such links can be understood as identity links in the sense of the Semantic Web, i.e., as 'same as' or 'very similar to', if we view both the megamodel and the destination as an ontology. Semantic annotation is enhanced by relying on ontological relationships between model elements such as the following:

- A system, an artifact, or a technology $x$ *uses* a concept or a language or a technology $y$. We offer the intuition that in such a case, one can somehow analyze $x$ to find evidence for the use of $y$. For instance, in the context of the *EMFModelAPI* module of Sec. 2.3, the generated model API uses (the concept of) 'tagging' such as 'generated' to tag generated Java members.
- A language, a technology, or a function $x$ *facilitates* a concept $y$. Without attempting a proper axiomatization, we offer the intuition that in such a case, one can use $x$ to achieve $y$. For instance, in the context of the *EMFModelAPI* module of Sec. 2.3, XMI could be said to facilitate (the concept of) 'persistence'.

**Realization** In MegaL, semantic annotation is also based on links, as in the case of 'Artifact binding'.

```
// module XML continued
XML = 'http://dbpedia.org/page/XML'
```

In MegaL, the following model elements can be semantically annotated:

- Concept entities; see the illustration below.
- Language entities, e.g., Java, XML, XSD, Ecore.
- Technology entities, e.g., JUnit, EMF, etc.
- Subtypes of the entity types Language, Technology, Concept. For instance, we could introduce a language subtype 'Markup language' and link it to the DBpedia resource of the same name; XMI should then be of type 'Markup language'.

Semantic annotation in MegaL is particularly potent when concepts are involved, as they are helpful in 'semantic tagging'. For instance, in the module *EMFModelAPI*, we may wish to convey that the API-defined functions for de-/serialization serve for persistence:

```
// module EMFModelAPI continued
Persistence : Concept
Persistence =
    'http://dbpedia.org/page/Persistence_(computer_science)'
CustomSerialize facilitates Persistence
CustomDeserialize facilitates Persistence
```

That is, the type *Concept* is used when declaring the entity *Persistence*. The concept is linked to an entity on DBpedia. The API-defined functions are declared to 'facilitate' persistence.

| |
|---|
| *xmlFile elementOf* XML: The analysis is reusable in that the load functionality of an XML API with well-formedness checking may be leveraged to accept or reject the file. |
| *javaFile elementOf* EcoreJava: Verification of the property of a given file to be an element of the Java subset targeted by the EMF code generator requires a custom analysis where different options may be considered: either the file is searched for a 'watermark' of EMF or the file is analyzed more deeply in terms of structure of the classes and the use of base types. |
| *xmlFile conformsTo xsdSchema*: Any sort of analysis for schema-/metamodel-like conformance should be reusable because conformance is usually an integral part of a technological (data) space. For instance, we may reuse any XSD schema validator to assert the validity of *xmlFile* to conform to *xsdSchema*. |
| *javaFile conformsTo EcoreJavaSpec*: We may use a domain-specific specification language for Java subsets so that we can capture the Java subset targeted by the EMF code generator or other Java subsets in a succinct manner. In this manner, we have established a new form of conformance, subject to a custom analysis, i.e., a custom interpreter. |
| *metaModel correspondsTo javaFiles*: We could attempt a reusable analysis by means of essentially running the suspected generator (*EMFGenerator*) on the metamodel and checking the result to be identical to the available grammar. However, it may be difficult to configure the generator in a way that it produces exactly the outcome. Thus, we may opt for a custom analysis: we either search for a watermark in the Java files which mentions the metamodel and the tool of interest or we check for the structural and nominal correspondence of classes and fields on the two sides. |

■ **Figure 6** Reusable versus custom analyses.

### 3.3 Pluggable analyses

**Problem** Megamodels may be easily checked in terms of well-formedness, i.e., correct use of the megamodeling vocabulary including proper declaration of referenced entities. However, verification of model versus system requires an evaluation of the relationships on actual artifacts. For instance, a conformance relationship necessitates a suitable conformance check. As noted before, in the descriptive mode, successful verification would mean that the derived model is a correct abstraction of the system; in the prescriptive mode, successful verification would mean that the derived system is complete and consistent with regard to the model.

We refer to the functionality for verification (evaluation) as *analyses*, as the related artifacts need to be analyzed (observed) to decide on the truth value of the declared relationship. The functionality may be readily available or it may need to be developed for the purpose of megamodeling. Thus, we distinguish *reusable* versus *custom analyses*. Fig. 6 discusses examples. Integration of reusable analyses and the development of custom analyses becomes an integral part of megamodeling, thus requiring infrastructural support.

**Solution** The megamodeling infrastructure provides a plugin architecture so that analyses for evaluating relationships are executed as part of automated verification. To this end, plugins are associated to relationship types. The execution of an anal-

## Interconnected Linguistic Architecture

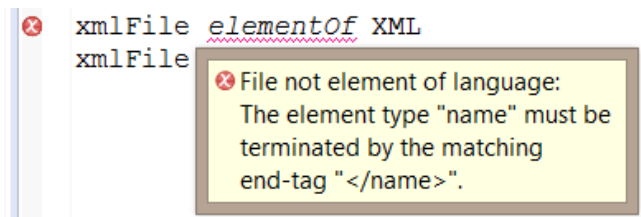

**Figure 7** A message fed back by the analysis for XML membership.

ysis provides messages and error feedback to users, thus aiding them in authoring megamodels (and instantiating them or deriving them by abstraction).

Plugins are preferred over other extension mechanisms, e.g., foreign language interface versus language embedding. An example of the former is ATL [16] where model transformations can apply helpers that are externally declared in Java classes. An example of the latter is Xbase [9] with its language inheritance for adding general-purpose language-like constructs to a newly modeled DSL. We advocate integration through plugins because of the importance of reuse. The aspects 'Artifact binding' and 'Model inference' also rely on a plugin infrastructure.

**Realization** MegaL/Xtext+IDE associates *plugins* for *analyses* with relationship types. MegaL/Xtext+IDE and plugins share a classpath. MegaL/Xtext+IDE uses bindings to resolve a JVM executable class file in the appropriate classpath. This method, combined with the incremental Java compiler provided by Eclipse, allows flexible modification of analysis behavior as needed.

For a given relationship, each associated plugin is invoked to check applicability. Verification of the megamodel against the system is considered incomplete, subject to appropriate warnings, if no plugin is applicable to a relationship. The analysis of each applicable plugin is executed and messages are displayed in the editor, as shown in Fig. 7. Absence of applicable plugins is also presented in the editor.

For instance, conformance in the XML story relies on a plugin like this:

```
class XMLConformsToXSD extends MegaLEvaluator {
  // Returns an evaluation report on the model element
  protected Report<Void> evaluate(Relationship element) {
    // Use SAX for validation; translate exceptions to report
    …
  }
}
```

MegaL/Xtext+IDE's plugin infrastructure is partially reflective; consider this:

```
1 conformsTo < Artifact * Artifact // Relationship type per prelude
2 ConformsToEvaluator : Plugin // Root plugin for conformance
3 ConformsToEvaluator = "classpath:ConformsToEvaluator"
4 conformsTo evaluatedBy ConformsToEvaluator
5 XMLConformsToXSD : Plugin // XML/XSD conformance
6 XMLConformsToXSD = "classpath:XMLConformsToXSD"
7 XMLConformsToXSD partOf ConformsToEvaluator
8 SAX : Technology // Semantic annotation of plugin
9 SAX = 'http://dbpedia.org/page/Simple_API_for_XML'
10 XMLConformsToXSD uses SAX
```

Johannes Härtel, Lukas Härtel, Marcel Heinz, Ralf Lämmel, and Andrei Varanovich

Relations like conformance (line 1) and plugins for analyses (lines 2 and 5) are model elements. Plugins are bound to the classpath (lines 3 and 6). One of the benefits is that plugins can be composed in the scope of a module (line 7). Another benefit is that we can also model knowledge about the analyses. For instance, we use a semantic annotation to express that the plugin at hand uses the XML technology SAX (lines 8–10).

### 3.4 Modularized models

**Problem** As software systems use software languages and technologies in similar ways, their linguistic architecture will be similar. Technologies can be used in different ways and in different combinations. This clearly calls for modular megamodels. In a prescriptive context, modules assist in reusing existing components, maybe from a system that serves for demonstration. In a descriptive context, modularization serves reuse of megamodels. Modularity needs to be harmonized with the need for artifact binding and megamodel execution.

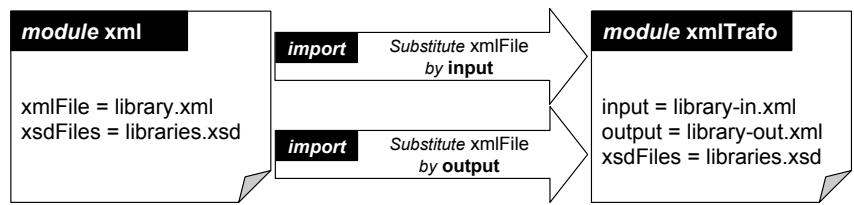

**Figure 8** Entity duplication by renaming (substitution) and rebinding along reuse of modules.

**Solution** The starting point is an import mechanism such that all declarations of an imported module are added to the importing module. Tool support for megamodeling shall take into account reuse, e.g., by providing designated IDE features. This includes quickly accessible documentation and comprehensive lists of locally declared versus imported model elements.

The import mechanism also facilitates *renaming* or, in fact, *substitution*. That is, when imported entities are renamed, distinct entities can be made the same.

Further, the import mechanism also facilitates *rebinding*. That is, entities bound in the imported module can be rebound in the importing module. Rebinding is needed because the imported module may have had artifact bindings to facilitate verification in a smaller context while the importing module requires different bindings as it is concerned with another context.

The import mechanism is also enhanced to facilitate *duplication* based on multiple imports of a given module. To this end, new names can be assigned to the duplicated entities.

A situation which combines renaming and duplication is shown in Fig. 8. The XML module is reused twice in the context of an assumed module for an XML transformation. We assume XML document and schema to be readily bound in the XML module. The entity for the XML document needs to be duplicated and rebound in each of

**Interconnected Linguistic Architecture**

the two instances so that we distinguish input and output documents of the XML transformation.

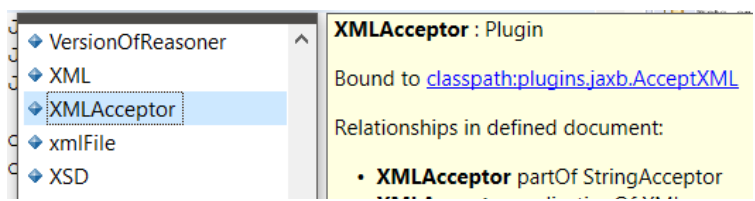

**Figure 9** Model element proposal and documentation in MegaL/Xtext+IDE.

**Realization**   MegaL's import mechanism has essentially copy-and-paste semantics modulo the following refinements: bindings in the importing module override bindings of the imported module; entities of the imported module can be renamed; if a module is imported more than once into a module (directly or indirectly), then non-renamed declarations are only imported once. In MegaL, we prefer renaming (substitution) along import over explicitly parameterized modules because it is hard to predict what megamodel entities would need to be parameters.

Rebinding affects verification of megamodel against the system. That is, the evaluation of a relationship from an imported module may need to be (re-)done for the importing module, if any of the related entities were rebound or a relevant pluggable analysis was only added by the importing module.

The MegaL/Xtext+IDE infrastructure supports modularity also at the IDE level. In particular, imported symbols are listed to assist in megamodel editing; see Fig. 9.

### 3.5 Transient artifacts

**Problem**   Many artifacts of a system's linguistic architecture are readily accessible in the system's source-code repository or in the post-build state on a file system. However, some artifacts related to a given scenario of language and technology usage may be missing in such a static view. For instance, scenarios related to web applications, web services, or web APIs involve (HTTP) requests and responses that are transient in nature. By this we mean that transient artifacts are only manifested temporarily as the result of specific actions performed by or with the system; also, transient artifacts are only accessible through means of interception. Transient artifacts may be merely transmitted as a network package or reside temporarily in computer memory; see Fig. 10 for an illustration. This is a challenge in megamodeling, if we assume that the data flow of the scenario should be modeled in a transparent manner.

**Solution**   Getting access to transient artifacts requires some form of amalgamation of system execution and megamodel execution. In the example of requests and responses, it requires mocking an incoming request and response interception—all triggered along the megamodel execution. For a modeler describing a system, transient artifacts might be hidden at first, but they are relevant to system understanding. Once transient artifacts are captured, they may be observed by analyses.

Johannes Härtel, Lukas Härtel, Marcel Heinz, Ralf Lämmel, and Andrei Varanovich

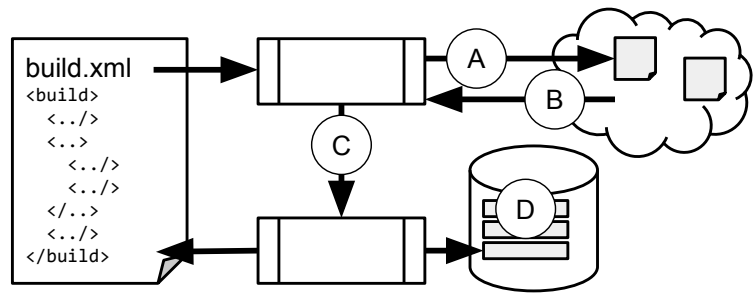

■ **Figure 10** A depiction of data flow and related transient states. A and B represent web request and response, respectively, C depicts piping of program output, and D shows transient data in memory or database.

More specifically, there are some options for access to transient artifacts: Firstly, interception at runtime (such as megamodel and system execution in the same JVM) may leverage a debugging interface or aspect-oriented programming. Secondly, interception in a network used for distribution may leverage network-snooping or the introduction of a middleman in a pipe. Thirdly, a system may also be instrumented to provide access to transient artifacts through accessor methods.

**Realization** MegaL/Xtext+IDE requires that transient artifacts are obtained during an inference phase (see Sec. 3.6) that precedes the evaluation of relationships. As of writing, we leverage the aforementioned option of code instrumentation so that transient artifacts are exposed by Java methods that execute the necessary steps and return the transient artifact as a Java object which is captured along megamodel execution. This option is suboptimal because it may imply some degree of system refactoring or extension.

For instance, the function *CustomSerialize* in the *API* story relies on a static method which reads an EMF resource with a model at the given URI:

```
class Persist {
  public static Resource deserializeXMI(URI model) { ... }
}
```

The function entity has to be bound to the method:

```
CustomDeserialize = 'classpath:.../Persist/deserializeXMI/1?p1=URI&r=Resource'
```

That is, the *deserializeXMI* method of class *Persist* is selected; the extra '/1' insists on methods with one parameter (in the view of overloading); the URL argument 'p1=...' describes the type of the first method parameter; likewise 'r=...' for the result type. Ontologically, this is a shortcut because we should say that the actual method *defines* a function and so we would have an artifact- (a fragment-) typed entity which is bound. For convenience's sake, the function entity in the megamodel gets directly bound instead.

**Interconnected Linguistic Architecture**

| |
|---|
| *Decomposition into parts*: The parthood-related structure of an artifact may be made systematically observable and amenable to part-level relationships; see also Sec. 3.8 on traceability. |
| *Artifact bindings* may follow a certain scheme such that they can be inferred. For instance, the names of languages and technologies, as used in the megamodels, may agree with names on knowledge resources such as DBpedia in a schematic manner, e.g., 'Java : Language' versus 'Java_(programming_language)' on DBpedia. |
| *Subset relationships between languages*: For instance, Java 7 is a subset of 8. These subset relationships should be interpreted transitively by inference. *'elementOf'* relationships should be inferred so that a statement of a file to be an element of Java 7 implies a statement of the same file to be an element of Java 8. The benefit of the inferred statements is that analyses for the different Java versions would be cross-validated. |

■ **Figure 11** Model inference scenarios.

### 3.6 Model inference

**Problem**   Some facts in the modeling process follow a pattern. Generally, descriptive as well as prescriptive modeling benefit from automating the inference of facts (model elements). Megamodeling would run into a scalability problem, if users were assumed to declare many routine elements. For instance, consider the notion of 'convention over configuration', e.g., in the context of web-application frameworks with certain conventions for folder or file names for model versus view versus controller. Fig. 11 lists further examples.

**Solution**   In the domain of ontologies and Semantic Web, the notion of inference addresses this problem [28]. A set of rules is applied to enrich a model or to support verification. This notion can be adopted to megamodels. In the domain of ontologies, there are some languages for rule-based deduction of knowledge, e.g., CONSTRUCT queries in SPARQL [19] or rules in F-Logic [1]. Declarative inference languages provide obvious benefits, but they cannot completely cover the needs of linguistic architecture. The aforementioned approaches work on model internal facts only, as opposed to inference and deduction on interconnected models, i.e., models connected to systems. Here, the content of a resource is taken into account and statements are retrieved from the subject's structure. In general, inference may need to incorporate methods for information retrieval.

**Realization**   MegaL/Xtext+IDE executes inference plugins repeatedly until a fixed point is reached. Inference plugins may only add megamodel elements, thereby making inference monotonic overall. For instance, the plugin for inferring parts of an Ecore-based metamodel takes this shape:

```
class EMFPartInferrer extends MegaLInferrer {
  // Returns an evaluation report and a model extension
  protected Report<Megamodel> infer(Entity element) { ... }
}
```

| |
|---|
| Navigation from a model element (an entity of type 'Artifact') to the actual (represented) artifact and vice versa. |
| Navigation from a compound artifact to fragments (parts) both on the side of the megamodel element and the actual artifact. (A view on the fragment may be based on an appropriate view of the compound artifact with the fragment being appropriately highlighted or unfolded.) |
| Navigation along trace links; see Sec. 3.8. |
| Navigation from a failed relationship or an error message or warning to the relevant analysis (plugin), involved artifacts, or relationship. |
| Navigation from a megamodel element (an entity or a relationship) to the relevant declaration or to other uses of the same name. (In this manner, one would also navigate, where necessary, to other modules.) |

■ **Figure 12** Navigation capabilities.

The model extension provided by EMFPartInferrer utilizes the reference implementation of the EMF: reading an Ecore-based metamodel is handled by the framework's resource API, the parsed metamodel elements are then translated into a format that our interpretation can understand. The plugin reuses domain-specific technology and is plugged in an entity type-driven manner:

```
Artifact evaluatedBy PartInferrer // Overall inferrer for parts
EMFPartInferrer partOf PartInferrer // Inferrer for EMF parts
```

### 3.7 Explorable connections

**Problem**  When a modeler or developer works with a megamodel, then errors may be returned as the result of executing analyses for the purpose of verification. The user may need to navigate through the megamodel and the system to better understand the error at hand. If such navigation is not supported by IDE-like capabilities, then the prescriptive and descriptive value of megamodels may be severely limited.

**Solution**  Megamodels must be explorable along all connections. This includes navigation from a megamodel element to related megamodel elements, system artifacts, plugins, and external resources. Fig. 12 lists navigation capabilities in detail.

Both prescriptive and descriptive modeling benefit from navigation; understanding a megamodel is alleviated by a unified interface for all sorts of connections.

The techniques that guide binding resolution can also be applied in navigation such that users can navigate to files, fragments, websites, etc. IDEs naturally allow opening many file types and APIs of editors or viewers can be leveraged to select a fragment in a file.

**Realization**  MegaL/Xtext+IDE leverages the editor and selection framework of Eclipse in combination with the URI artifact providers discussed earlier. Eclipse provides an API for opening editors and jumping to text ranges, which is paired with IDE specific,

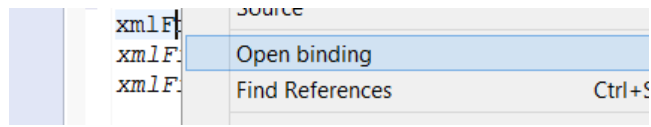

■ **Figure 13** The open binding IDE integration in MegaL/Xtext+IDE, instead of going to the model element's declaration, the editor will navigate to the bound artifact.

AST-navigating artifact providers. Users may navigate by invoking that functionality from a context-sensitive menu; see Fig. 13 for an illustration.

## 3.8 Traceability links

**Problem** When code generators, mapping technologies, translators, and yet other forms of software transformations or analyses are used in a software system, then their usage implies structured relationships between artifacts. For instance, mapping an object model to a database schema implies a systematic correspondence between classes versus tables and within each class/table pair another level of correspondence between fields and columns. Such structural decomposition of relationships is also possible for other relationships—notably for conformance.

More generally, relationships between artifacts that consist of parts may be represented as collections of traceability links that serve understanding of transformations [26] and other relationships. Unless the technology of interest has readily left suitable traces, traceability links need to be inferred and made amenable to navigation.

**Solution** The starting point is access to the parthood-related structure of artifacts, as discussed in the context of model inference (Sec. 3.6). In a next phase, traceability links are to be inferred from (the parts of) related artifacts. To this end, the corresponding relationship types are associated with custom functionality. Conceptually, a trace (a collection of traceability links) is a bipartite graph. Technically, the graph can be directly expressed as a model extension (again, in the sense of model inference) with the relationship of interest applied to the parts of the compound artifacts. The artifact binding for parts immediately enables navigation.

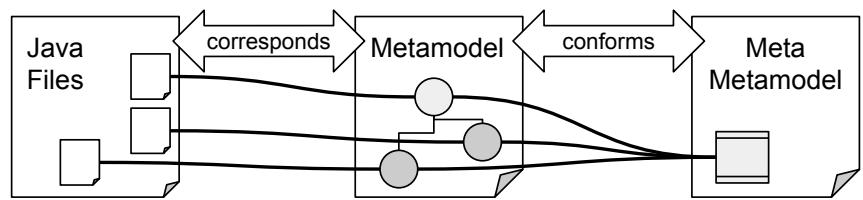

■ **Figure 14** After code generation (illustrated in section 2), the Java files depicted left correspond to the originating metamodel, which in turn conforms to the metametamodel.

**Realization** Traces are recovered after the inference phase. For some given correspondence fact, the trace is deducted and published to the model by listing the fragments

| | |
|---|---|
| ⌄ xsdFiles | javaFiles |
| /xs:schema/xs:complexType | org/softlang/company/xjc/Employee.java |
| /xs:schema/xs:element#0 | org/softlang/company/xjc/Company.java |
| /xs:schema/xs:element#1 | org/softlang/company/xjc/Department.java |
| ⌄ xmlFile | objectGraph |
| ⌄ company/department#0 | org.softlang.company.xjc.Department@5fd1a6aa |
| ⌄ employee#0 | org.softlang.company.xjc.Employee@1a56a6c6 |
| address:Utrecht | Utrecht |
| name:Erik | Erik |
| salary:12345 | 12345.0 |
| › employee#1 | org.softlang.company.xjc.Employee@748e432b |

■ **Figure 15** Explorable trace links in MegaL/Xtext+IDE for an extended XML story with involvement of XML-data binding, i.e., Java-class generation from an XML schema. The trace at the top shows similarity of XSD schema versus Java classes. The trace below shows similarity of XML document versus Java object (past deserialization). The indented rows are fragments (part of) the files. Fragmented URIs are used where applicable. Similar traces arise in the EMF story with generation and serialization of Sec. 2.

and their correspondence in the form of dedicated model elements and interconnections. See Fig. 14 for an illustration and Fig. 15 for a snapshot of the realized trace visualization plugin. The realization shows all correspondence relations between entities, i.e., the trace, as rows. The nesting structure of the rows that is used for collapsing follows the actual composition of the corresponding entities. Since the congruence of structure on both sides of the trace is not necessarily given (e.g., by having intermediate fragments or varying nesting), the realization determines one root on the left column and traverses parts recursively. Thereby, the rows are populated and the nesting is uniquely defined.

## 4  Evaluation

Our evaluation is twofold. Firstly, we survey literature in the context of megamodeling with regard to the aspects of interconnected megamodels. Secondly, our implementation and our case studies are assessed on the aspects.

### 4.1  Literature survey

We searched for conference and journal publications on DBLP with mention of megamodel(s) and megamodeling. In this manner, we located papers that enhance the megamodeling notion [7, 12], identify application domains [30, 15, 11, 17], and consolidate the foundation [25, 3, 4, 10].

In these papers, we aimed to identify occurrences of our aspects of interconnected megamodels. The results are given in the upper part of Tab. 2 where the size of a bullet corresponds to the level of coverage: an empty cell means that the relevant

**Interconnected Linguistic Architecture**

| | | | | | | | | |
|---|---|---|---|---|---|---|---|---|
| [30] | | | · | | • | | • | · |
| [7] | | • | · | | • | | · | |
| [25] | | • | • | • | • | | · | |
| [15] | | · | | | · | | · | |
| [11] | | • | · | | • | | | |
| [17] | | • | · | • | | | · | |
| [12] | | • | · | • | | · | | |
| [3] | | • | · | • | | · | | |
| [4] | | • | • | • | | · | · | |
| ([10]) | | • | • | · | · | • | | • |
| $L_3$ | | ◯ | | ◯ | | | ◯ | |
| $L_2$ | | | ⊗ | × | × | × | | ◯ |
| $L_1$ | | ⊗ | × | | ◯ | ◯ | × | × |
| | 3.8 | 3.1 | 3.6 | 3.3 | 3.7 | 3.4 | 3.2 | 3.5 |
| | Traceability links | Artifact binding | Model inference | Pluggable analyses | Explorable connections | Modularized models | Semantic annotations | Transient artifacts |

■ **Table 2** Upper part: Mapping of other megamodeling related papers to our aspects, bigger dots depict stronger focus. The paper [10] is shown in parentheses because of an overlap of the authors with those of the current paper. Lower part: maturity of MegaL/Xtext+IDE, regarding implementation (◯) and the demonstration (×) based on maturity levels—ranging lowest to highest from $L_1$ to $L_3$.

aspect was not present in the paper; a small bullet depicts some coverage through the identification of the topic or some limited implementation; a big bullet depicts full coverage and extension.

It turns out that artifact binding, traceability links, and inference are covered very well. Our work brings these aspects to the area of linguistic architecture—as a particular form of megamodeling. Some inspected approaches use model transformation traces (e.g., for impact analysis); this requires identifying source and target elements of a transformation [11, 17, 3, 4, 10]. We automate this approach based on the aspects 'Model inference' and (recovery of) 'Traceability links'. Analysis of models (in the sense of 'pluggable analyses'), is not covered as well. Analyses are mainly delegated to making the implicit structure explicit and checking model constraints thereafter. Some approaches support such checks by native code, in a 'pluggable analysis' fashion [11, 17], others rely on established model checking solutions which utilize a rule specification format [30, 7, 25, 15, 17, 12]. Exploration is covered in several cases [10, 25, 15, 11, 17, 3], mainly due to visualization of complex inter-model connections such as traceability links. Modularization is not widely present. Our notion of semantic annotations as well as transient artifacts are least covered by other approaches. Even though linking facilities are utilized often [30, 15, 11, 17, 4]; these facilities are limited in terms of the kind of artifacts they can address. Transient artifacts are only covered marginally.
3-22

One publication suggested using a general purpose programming language as the foundation for model transformation [30] such that one may imagine to bind transient artifacts to the model.

### 4.2 Status of the realization

MegaL/Xtext+IDE covers the different aspects of interconnected megamodels at different levels of maturity. We distinguish between implementation, i.e., language workbench and megamodel execution versus demonstration of an aspect in terms of a comprehensive case study. Our findings are listed at the bottom of Tab. 2. The lowest level ($L_1$) in the table corresponds to implementation or demonstration outlines, the highest denote aspects we cover comprehensively. (We note that the implementation of 'Semantic annotations' is identical with the one of 'Artifact binding'.)

We mention that the most advanced case study is on XML-data binding with XML. There are several smaller case studies, e.g., related to language implementation with ANTLR. The work on megamodels concerning EMF, as shown to some extent in the present paper, is ongoing.

Our understanding of instances, where implementation and demonstration diverge, is the following: (i) when the former outranks the latter, then our solution is highly potent, and the possibilities are not exhausted in the case studies, (ii) when the inverse situation happens, then an aspect's demonstration required fine-tuning using implementation insight, surpassing the current language tool set. For instance, exploration is sketched with usage of some Eclipse functions that are specific to a case study's domain. The ideal infrastructure should be more generic.

## 5 More related work

**Ontology engineering** Our story-based approach of Sec. 2 is inspired by Presutti et al.'s task-oriented process for ontology engineering [22] which leverages customer knowledge, using brief stories from a problem domain to obtain so called *competency questions*, which are used to incrementally refine an ontology.

Our modular approach to megamodeling is similar to Ruy et al.'s reuse of ontological patterns [24] catering for three levels of granularity: foundational versus core versus domain. Likewise, we have three levels: the prelude as foundation of megamodeling, technology megamodels as core, further instantiations and composition to apply megamodels to concrete systems. Modularization is also addressed by Grau et al. [13] in the sense of module extraction, where the resulting ontology fragments cover a topic (a set of terms) completely; reuse is, in their case, alleviated.

Semantic annotations and ontologies in the sense of the Semantic Web also finds other applications in model-driven and software language engineering. In [6], annotations guide a software reverse engineering process. In [29], ontologies are used to analyze software, by detecting 'smells' of anti-patterns. In [23], ontological information is extracted from API signatures.

**Traceability**   This is an issue in several areas of model-driven and software language engineering [8, 21, 14, 5]. For instance, in [5], the authors suggest domain-specific traces, which cover the execution of DSLs. In [14], automatic generation of traces is elaborated. In order to detect interconnection, an incremental query engine detects *soft links*, rather than explicit cross model references.

**Megamodeling infrastructure**   Our approach is related to Melanie [2]. Melanie supports linguistic and ontological instantiation, features both textual and graphical editing, allows third-party extensions, supports fact inference and analysis. Our approach additionally involves introspection of bound artifacts along analysis (verification) and inference. Hilliard et al. [15] describe a system which also deals with model relationships and checking their validity using OCL and the AM3 engine. Our approach provides pluggability and technological space independence, thereby enabling essential reuse of analyses for relationships. Other work on megamodeling also does not generally apply to usage scenarios of technologies across spaces. Jouault et al. [17] address the problem of DSL integration based on megamodeling in a MDE context. *MoScript* [18] uses megamodeling for scripting in model repositories. Seibel et al. [27] use dynamic megamodels for managing runtime maintenance tasks on given models.

## 6   Concluding remarks

We have presented the notion of interconnected linguistic architecture with the objective of making megamodels more useful, specifically for the documentation of software technologies and understanding their usage in software systems. The focus is on connecting model and system in a transparent, automated, modularized, and explorable manner. Based on several scenarios ('stories') in the broader scope of the Eclipse Modeling Framework (EMF), we captured linguistic architecture of language use and definition, model-to-model transformation, code generation, and runtime serialization. We identified recurring problems in linguistic architecture modeling which limit developers and modelers when aiming at abstraction, instantiation, alignment, and verification, as part of prescriptive and descriptive megamodeling.

   We addressed the identified problems by identifying fundamental aspects of connection across megamodels, systems, and developer experience. The notion of interconnected linguistic architecture is realized in the language MegaL and the accompanying infrastructure MegaL/Xtext+IDE. We also carried out a literature study on megamodeling which shows the scattered presence of isolated aspects of interconnected megamodels. Thereby, our primary research contribution is the proper identification and integration of the different aspects of interconnected linguistic architecture. We believe that the identified aspects are general in nature and useful for technology documentation overall; they may be considered 'principles' of interconnected linguistic architecture.

   Future work should address the completeness problem of a megamodel: When does it capture enough (all) important properties of technology usage in all typical scenarios? On a related account, one should also study this question: How to systematically derive megamodels from classic documentation and also how to perform alignment or

**Johannes Härtel, Lukas Härtel, Marcel Heinz, Ralf Lämmel, and Andrei Varanovich**

even megamodel 'inference' more automatically? Another related problem is the use of appropriate vocabularies in megamodels. For instance, we may need to discover more or less complex forms of synonyms such as 'metamodel and model' versus 'model and instance model'. A solution requires IR techniques and ontology engineering.

Further future work concerns improved realizations of some of the aspects. (i) We have noticed that the semantics of multiple imports with renaming and rebinding is complex and deserves proper formalization on its own. (ii) Plugins should be modeled more strongly within the megamodel so that the purpose and the characteristics of each plugin or part thereof can be observed without consulting the plugin's implementation. (iii) Interception of transient artifacts of diverse kinds should be systematically researched as opposed to our dependence on code instrumentation. (iv) Inference should rely on enough declarative properties and suitable analyses to guarantee termination and determinism. (v) The understanding of the *time* dimension in linguistic architecture needs to be improved, with a focus on process, build cycles, and evolution of state in programs or databases.